\documentclass[11pt]{iopart}
\usepackage[british]{babel}
\usepackage{color}
\usepackage{mathrsfs}
\usepackage{setstack}
\usepackage{iopams}
\usepackage[T1]{fontenc}
\usepackage{graphicx}
\usepackage{color}
\usepackage{enumitem}
\renewcommand{\text}[1]{\mathrm{#1}}

\usepackage{hyperref}

\usepackage{cite}

\hypersetup{
  colorlinks=true,
citecolor=black,
linkcolor=black,
urlcolor=black
  }

\usepackage[colorinlistoftodos,prependcaption,textsize=small]{todonotes}

\begin{document}

\title{Deterministic and randomized motions in single-well potentials}

\author{Micha{\l} Mandrysz$^\dagger$ and Bart{\l}omiej Dybiec$^\dagger$}

\address{$\dagger$ Marian Smoluchowski Institute of Physics,
and Mark Kac Center for Complex Systems Research, Jagiellonian University, ul.
St. {\L}ojasiewicza 11, 30--348 Krak\'ow, Poland}
\ead{michal.mandrysz@student.uj.edu.pl,bartek@th.if.uj.edu.pl}

% \date{\today}

\begin{abstract}
Newtonian, undamped motion in single-well potentials belong to a class of well-studied conservative systems.
Here, we investigate and compare long-time properties of fully deterministic motions in single-well potentials with analogous randomized systems.
We consider a special type of energy-conserving randomization process: the deterministic motion is interrupted by hard velocity reversals $\vec{v}(t_i)\to-\vec{v}(t_i)$ at random time instants $t_i$.
In the 1D case, for fixed initial conditions, the differences in probability distributions disappear in the long-time limit making asymptotic densities insensitive to the selection of random time instants when velocity is reversed. Substantially different probability distributions can be obtained, for instance, through the additional randomization of initial conditions.
Analogously, in 2D setups, the probability distributions asymptotically are insensitive to velocity reversals.
\end{abstract}


\maketitle

%%%%%%%%%%%%%%%%%%%%%%%%%%%%%%%%%%%%%%%%%%%%%%%%%%
%                                                %
%    BEGINNING OF TEXT                           %
%                                                %
%%%%%%%%%%%%%%%%%%%%%%%%%%%%%%%%%%%%%%%%%%%%%%%%%%

%%%%%%%%%%%%%%%%%%%%%%%%%%%%%%%%%%%%%%%%%%%%%%%%%%%%%%%%%%%%%%%%%%%%%%%%%%%%%%%%%%%%%%%%%
%%%%%%%%%%%%%%%%%%%%%%%%%%%%%%%%%%%%%%%%%%%%%%%%%%%%%%%%%%%%%%%%%%%%%%%%%%%%%%%%%%%%%%%%%
%%
%% introduction
%%
\section{Introduction and motivation\label{sec:introduction}}

Newtonian dynamics provides an accurate, deterministic description of macroscopic systems.
The prerequisite for its application is the knowledge of the initial conditions. In the absence of dissipation, the Newtonian dynamics in static potentials is  conservative, i.e., mechanical energy is constant.
Consequently, the motion in the phase space continues along a constant energy curve.
Variable, stochastic environment can be introduced by time-dependent parameters or by the introduction of noise \cite{gammaitoni2009,anishchenko1999}.
Noise is usually considered to be an approximate description of interactions between the studied particle/body and its environment.
The action of additive noise, which is not counterbalanced by some other mechanism, e.g., dissipation, breaks both mean and instantaneous energy conservation \cite{mallick2007anomalous,sekimoto2010stochastic}.
Nevertheless, a special class of disturbances in dynamical systems can still preserve the energy of the system, see Dybiec et al. \cite{dybiec2018conservative}.
For instance,  the disturbance which reverses the velocity vector at random instants of time does not break the energy conservation at all.
Presence of the stochastic component in the system dynamics transforms state variables into random variables.
Properties of perturbed systems can be studied by means of statistical physics.
Analogously, fully deterministic systems can also be studied via notions of probability theory.
The long observation of deterministic trajectories allows one to estimate the probability density of finding a system in a given state \cite{walters2000introduction}.

Random perturbations in dynamical systems play a crucial role in noise-driven systems.
Among noise-induced effects there are well-known and celebrated effects of stochastic resonance \cite{anishchenko1999,gammaitoni1998}, resonant activation \cite{doering1992,pankratov2000,iwaniszewski2008resonant}, stochastic synchronization \cite{anishchenko1997} and direct transport (ratcheting effect) \cite{reimann2002}.
The presence of noise can also induce dynamical multimodality \cite{dybiec2007c,iwaniszewski2008transient,calisto2017forced} and dynamical hysteresis \cite{mahato1994,berglund2002hysteresis}.
All these effects originate in or owe its efficiency due to the combined action of deterministic and random forces.
For instance, in the stochastic resonance, the combined action of the deterministic perturbation and fine-tuned noise amplify the system response by making the random process of passages over the potential barrier optimally synchronized with the periodic modulation of the potential barrier.

Current investigations are motivated by the L\'evy walk model \cite{zaburdaev2015levy}.
The L\'evy walk is a dynamical process with simple spatiotemporal coupling.
The random walker moves with a constant velocity $v$ for a random independent and identically distributed times $\tau_i$ ($\tau_i>0$) followed by a velocity flip, i.e., $v\to -v$.
The time $\tau$ determines the flight time and the distance traveled in between velocity reversals.
The mechanical energy of the L\'evy walker is constant and equal to kinetic energy.
L\'evy walks were successfully applied to a plenitude of models including, among others, dynamics of tracer particles in weakly
chaotic systems \cite{zumofen1993power,geisel1984anomalous,solomon1993}, cold atom systems \cite{kessler2012theory}, random search
strategies \cite{shlesinger1986,lomholt2008levy} and intracellular motion \cite{song2018neuronal,chen2015memoryless}.

In Dybiec et al. \cite{dybiec2018conservative} the extension of the L\'evy walk scenario including deterministic forces has been suggested.
For the deterministic motion in a single-well potential velocity continuously changes due to the deterministic $-V'(x)$ force.
Therefore, in 1D, the absolute value of velocity $|v(t)|$ decreases as the particle moves towards large $|x|$ and at large $|x|$ the motion is smoothly reversed when $|v(t)|=0$.
In addition to deterministic soft velocity reversals, hard velocity reversals at random time instants $t_i$ have been introduced.
Hard velocity reversals are responsible for velocity flips $\vec{v}(t_i)\to -\vec{v}(t_i)$.

Here, we summarize some of the findings and further extend the model of Ref.~\cite{dybiec2018conservative}.
Using the specific type of a multiplicative disturbance which reverses the velocity vector at random instants of time the deterministic system is randomized.
The process of velocity reversals can be also interpreted as a dichotomous (assuming two values $\pm 1$), multiplicative perturbation $\eta(t)$ to the velocity $\vec{v}(t) \to \eta(t)\vec{v}(t)$.
On the one hand, this process does not break the energy conservation.
On the other hand, it introduces a discontinuity in the phase space because of discontinuous velocity changes.
Nevertheless, the system stays on the same orbit, and asymptotically its statistical properties are the same as properties derived from the very long observation of the fully deterministic motion.
Significant changes in its statistical properties are observable due to a different type of randomizations, e.g., randomization in its initial conditions. Randomization of initial conditions makes the total energy random variable and produce densities which significantly deviate from $u$-shaped densities appearing for fixed initial conditions accompanied with hard velocity reversals \cite{dybiec2018conservative}.
Within the studied model, the main source of the randomization is due to hard velocity reversals. Therefore, the model resembles the model of hard spheres \cite{scalas2015velocity} in the square-well potential which in 2D, two-particle,  molecular dynamics ensemble is able to produce similar probability density functions for the velocity like the model studied within the current manuscript.
Nevertheless, both models bear fundamental differences:
the setup studied here is a single particle and non-zero potential model of randomized motion
while the one considered in \cite{scalas2015velocity} is a multi-particle 2D and 3D model of the hard sphere gas.

The model under consideration is presented in the next section (Sec.~\ref{sec:model}).
Section Results (Sec.~\ref{sec:results}) presents the outcome of our analysis.
Starting from  the conclusions from Dybiec et. al.~\cite{dybiec2018conservative} about hard velocity reversals we provide universal, analytical derivation of general formulas which, for simplest cases of single-well potentials, were constructed semi-analytically and numerically  in \cite{dybiec2018conservative}. Next, we ask an ancillary question about the long-time behavior of a class of fully deterministic systems in which initial conditions are randomized (Section~\ref{sec:1d}).
Finally, the higher dimensional case (2D) of quasi-periodic dynamics in 2D single-well potentials is analyzed with and without hard velocity reversals (Section~\ref{sec:2d}).
The manuscript is closed with Summary and Conclusions (Sec.~\ref{sec:summary}).
More technical derivations are moved into the appendices (\ref{app:derivation-p} -- \ref{app:derivation-spherical}).

%%%%%%%%%%%%%%%%%%%%%%%%%%%%%%%%%%%%%%%%%%%%%%%%%%%%%%%%%%%%%%%%%%%%%%%%%%%%%%%%%%%%%%%%%
%%%%%%%%%%%%%%%%%%%%%%%%%%%%%%%%%%%%%%%%%%%%%%%%%%%%%%%%%%%%%%%%%%%%%%%%%%%%%%%%%%%%%%%%%
%%
%% model
%%
\section{Model\label{sec:model}}

\begin{figure}
    \centering
    \includegraphics[width=0.85\columnwidth]{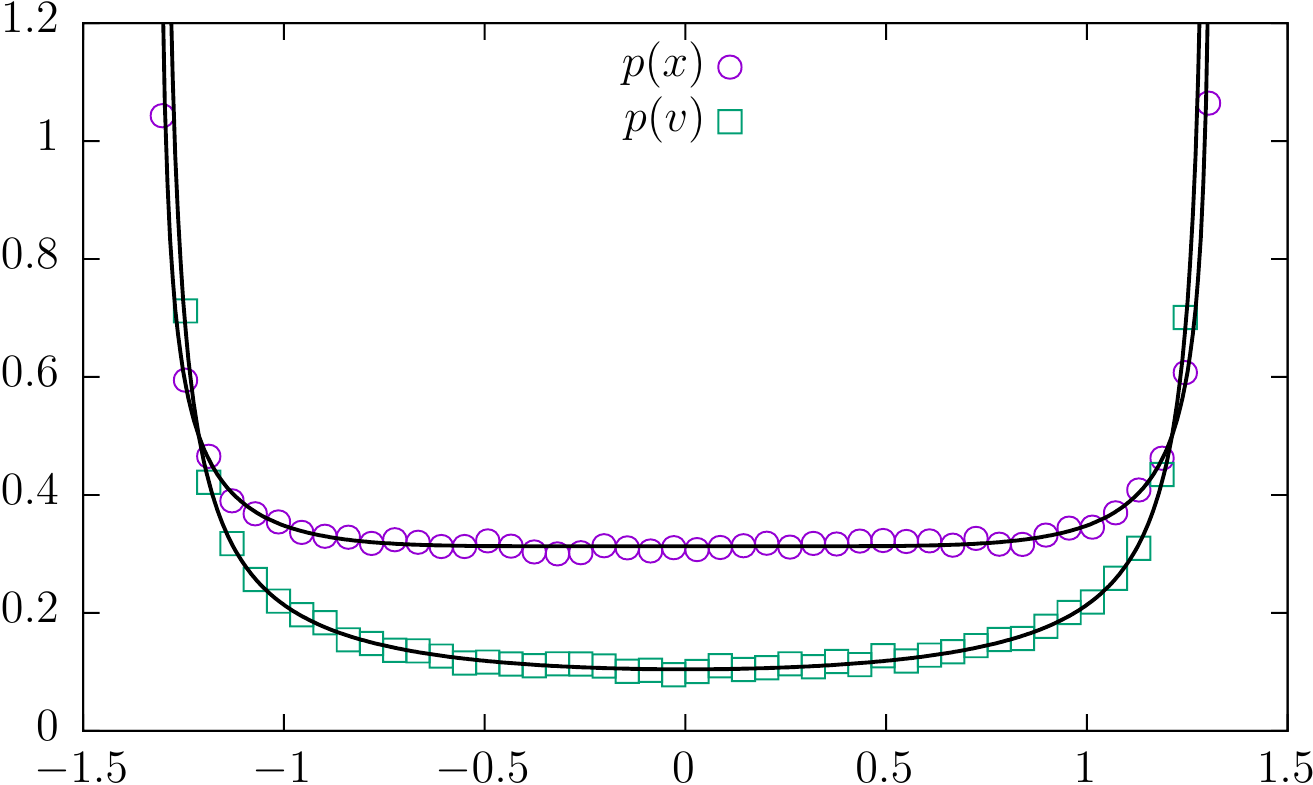}
        \caption{Stationary densities $p(x)$ and $p(v)$ for $n=6$ with $\kappa=1/6$, see Eq.~(\ref{eq:potential}) and \cite{dybiec2018conservative}. Solid lines present exact results, see Eqs.~(\ref{eq:px}) and~(\ref{eq:pv}), while points correspond to results of stochastic simulations of the model with velocity reversals.}
    \label{fig:fixed-n6}
\end{figure}

The Newton's equation
\begin{equation}
    m \frac{d^2 x(t)}{dt^2}=-V'(x)
    \label{eq:newton1d}
\end{equation}
with
\begin{equation}
    V(x)=\kappa |x|^n\;\;\;\;\;\;\;\;\;\;\;\;(\kappa>0,n>0)
    \label{eq:potential}
\end{equation}
describes a periodic motion \cite{landau1988theoretical} with the period $T$ given by
\begin{equation}
    T=\frac{2}{n}\sqrt{\frac{2\pi m}{E}}\left[ \frac{E}{\kappa} \right]^{\frac{1}{n}} \frac{\Gamma\left( \frac{1}{n} \right)}{\Gamma\left( \frac{1}{2} + \frac{1}{n} \right)},
    \label{eq:period}
\end{equation}
where $\Gamma(\dots)$ is the Euler Gamma function.
Due to the absence of damping the system described by Eq.~(\ref{eq:newton1d}) is conservative.
Its total energy $E$ is equal to
\begin{equation}
    E= \frac{1}{2}mv^2+V(x)= \frac{1}{2}mv^2+\kappa|x|^n.
\end{equation}
From a single trajectory, after a long observation time, it is possible to calculate the probability of observing the particle in the vicinity of the point $x$
\begin{equation}
    p(x)= \frac{2}{T} \left[ \frac{2}{m} \left( E - \kappa|x|^n \right) \right]^{-\frac{1}{2}}.
    \label{eq:px}
\end{equation}
Analogously, the probability of recording velocity $v=\dot{x}$ reads
\begin{equation}
    p(v)=\frac{2m}{T\kappa n} \left[ \frac{1}{\kappa } \left( E - \frac{1}{2}m v^2 \right)  \right]^{\frac{1}{n}-1}.
    \label{eq:pv}
\end{equation}
In Eqs.~(\ref{eq:px}) and~(\ref{eq:pv}) $T$ stands for the period of the motion which is given by Eq.~(\ref{eq:period}).
The total energy $E$ is fixed by the initial condition.
Densities given by Eqs.~(\ref{eq:px}) and~(\ref{eq:pv}) are similar to the arcsine distribution with modes located at $V(x)=E$ for $p(x)$ or $mv^2/2=E$ for $p(v)$, see Fig.~\ref{fig:fixed-n6}.
For selected values of $n$ Eqs.~(\ref{eq:px}) and (\ref{eq:pv}) were derived in \cite{dybiec2018conservative} using semi-analytical and numerical methods.
The detailed, universal, analytical derivation of general formulas~(\ref{eq:px}) and~(\ref{eq:pv}) is presented in~\ref{app:derivation-p}.

In Dybiec et al. \cite{dybiec2018conservative}, the 1D deterministic motion governed by Eq.~(\ref{eq:newton1d}) with fixed initial condition is perturbed by hard velocity reversals which do not break the energy conservation.
Such hard velocity reversals at random time instants are separated by a random time $\tau$.
More precisely, the sequence of waiting times $\tau_i$ follows a one-sided probability density $p(\tau)$.
The particle moves according to Eq.~(\ref{eq:newton1d}) as long as $t\leqslant \tau_1$.
Then, the velocity is reversed $v(\tau_1)\to -v(\tau_1)$ and the particle moves deterministically till the next velocity reversal, i.e., while $\tau_1 < t \leqslant \tau_1+\tau_2$.
The next velocity reversal is at time $\tau_1+\tau_2$, i.e., $v(\tau_1+\tau_2)\to -v(\tau_1+\tau_2)$.
The whole procedure is repeated iteratively.
At time instants $\tau_1$, $\tau_1+\tau_2,\dots$ only the velocity is reversed while the position is unaffected.
Therefore, we solve Eq.~(\ref{eq:newton1d}) in time intervals $0\leqslant t \leqslant \tau_1, \tau_1 \leqslant t \leqslant \tau_1+\tau_2, \dots $ with the initial conditions determined by the final position and the reversed final velocity from a preceding interval.

%%%%%%%%%%%%%%%%%%%%%%%%%%%%%%%%%%%%%%%%%%%%%%%%%%%%%%%%%%%%%%%%%%%%%%%%%%%%%%%%%%%%%%%%%
%%%%%%%%%%%%%%%%%%%%%%%%%%%%%%%%%%%%%%%%%%%%%%%%%%%%%%%%%%%%%%%%%%%%%%%%%%%%%%%%%%%%%%%%%
%%
%% model
%%
\section{Results \label{sec:results}}

First, we reexamine the periodic motion in 1D single-well potentials (Sec.~\ref{sec:1d}).
Next, we switch to the 2D single-well potentials (Sec.~\ref{sec:2d}).
We study similarities and differences between 1D and 2D models with particular attention to the consequences of the fact that orbits in 2D single-well potentials do not need to be closed.

%%%%%%%%%%%%%%%%%%%%%%%%%%%%%%%%%%%%%%%%%%%%%%%%%%%%%%%%%%%%%%%%%%%%%%%%%%%%%%%%%%%%%%%%%
\subsection{1D randomized motions \label{sec:1d}}

As it was shown in Dybiec et al. \cite{dybiec2018conservative} asymptotic $p(v)$ and $p(x)$ densities are robust to the exact shape of $p(\tau)$.
At the same time, transient behavior can display some sensitivity to the $p(\tau)$.
At short times randomized motion following Eq.~(\ref{eq:newton1d}) produces stationary states which can be decorated by fingerprints of initial conditions.
This effect is the most persistent for heavy-tailed distributions of reversal times which are characterized by the diverging mean.
Regardless of the waiting time distribution $p(\tau)$ peaks corresponding to the initial condition decay over time.

Asymptotically, hard velocity reversals at random time instants produce the same position and velocity distributions regardless of $p(\tau)$, i.e., asymptotics densities are given by Eqs.~(\ref{eq:px}) and (\ref{eq:pv}) with appropriately adjusted parameters.
Consequently, we now have three scenarios which end up with the same $p(x)$ and $p(v)$ distributions, namely when we build those distributions out of:
\begin{enumerate}[label=(\roman*)]
    \item{very long observation of a single deterministic trajectory with a fixed initial condition,\label{itm:1}}
    \item{ensemble averaging of final velocities and positions with random initial conditions uniformly distributed over the fixed energy orbit,\label{itm:2}}
    \item{ensemble averaging of final velocities and positions with fixed initial condition accompanied with random velocity reversals.\label{itm:3}}
\end{enumerate}

The scenario \ref{itm:1} corresponds to the time averaging over a single very long trajectory, which evolves according to the Newton equation of motion.
For the very long observation time, with the fixed frequency, we sample $x$ and $v$ which are used to estimate $p(x)$ and $p(v)$ densities.
The protocol \ref{itm:2} is the ensemble averaging of trajectories with the fixed total energy $E$ but different initial velocities and positions.
More precisely, in \ref{itm:2} we generate a large number of starting points (initial conditions) which are uniform on the fixed energy curve. For each initial condition we evolve $x(t)$ and $v(t)$ according to the Newton equation. From the ensemble of final points we estimate $p(x)$ and $p(v)$ densities.
Finally, the scenario \ref{itm:3} assumes randomized motion with a fixed initial condition.
In \ref{itm:3} we have a fixed initial condition but the velocity is randomized: at random time instants $v(t_i)\to-v(t_i)$. For~\ref{itm:3}, we see transient fingerprints of the initial condition which are sensitive to the  waiting time distribution $p(\tau)$. At the same time, asymptotics $p(x)$ and $p(v)$ densities are robust with respect to the initial conditions on the constant energy curve and the waiting time density $p(\tau)$, but they depend on the total energy $E$.

Scenarios \ref{itm:1} -- \ref{itm:3} produce the same $p(x)$ and $p(v)$ distributions under the condition that the initial energy is the same, the observation time is long
enough to reach the asymptotic regime.
Exemplary Fig.~\ref{fig:fixed-n6} demonstrates the perfect agreement between results of numerical simulations of the scenario \ref{itm:3} for $V(x)=x^6/6$ with stationary densities $p(x)$ and $p(v)$ given by Eqs.~(\ref{eq:px}) and~(\ref{eq:pv}) which are derived using scenario \ref{itm:1}.
Therefore, in addition to decaying deterministic peaks, the main consequence of such randomization is the equivalence of ensemble and time averaging of motions starting with the fixed initial condition.
This equivalence allows one to construct the exact distributions, see Eqs.~(\ref{eq:px}) and~(\ref{eq:pv}).

Deviations from the obtained probability distributions can be easily produced by the addition of noise to Eq.~(\ref{eq:newton1d}).
Presence of white Gaussian, Ornstein-Uhlenbeck or Markovian dichotomous noise results in the pumping of the energy to the system \cite{mandrysz2019energetics-pre,albeverio1994long,albeverio2000longtime}.
Consequently, the system's trajectories in the phase space are dispersed and do not stay on closed orbits.
Nevertheless, the trajectory $x(t)$ crosses $x=0$ infinitely many times \cite{mao2007stochastic}.
The addition of noise without the introduction of the counterbalancing dissipative term introduces significant disturbances to the model.
The alternative, more gentle scenario is to consider random initial conditions.
In such a case, contrary to the fixed initial condition, the motion is not restricted to the single orbit, but a set of orbits determined by random initial conditions.
After fixing the initial energy, the motion with hard velocity reversals is continued along the constant energy curve as in the scenario \ref{itm:3}.
For random initial conditions, stationary densities can be calculated by the composition of probability densities
\begin{equation}
    p(x)=\int p(x|\lambda)p(\lambda)d\lambda
    \label{eq:comp-x}
\end{equation}
and
\begin{equation}
    p(v)=\int p(v|\lambda)p(\lambda)d\lambda,
    \label{eq:comp-v}
\end{equation}
where $\lambda$ indicates a randomized parameter in $p(x)$ and $p(v)$ densities, while $p(\lambda)$ is the probability density of the randomized parameter.
For simple distributions $p(\lambda)$ resulting densities given by Eqs.~(\ref{eq:comp-x}) and~(\ref{eq:comp-v}) can be calculated exactly.

\begin{figure}
    \centering
%     \begin{tabular}{cc}
%     \includegraphics[width=0.48\columnwidth]{ric-v-n2-02-11} & \includegraphics[width=0.48\columnwidth]{ric-v-n2-09-11}\\
%     \includegraphics[width=0.48\columnwidth]{ric-v-n6-02-11} &
%     \includegraphics[width=0.48\columnwidth]{ric-v-n6-09-11}    \\
%     \end{tabular}
    \includegraphics[width=0.98\columnwidth]{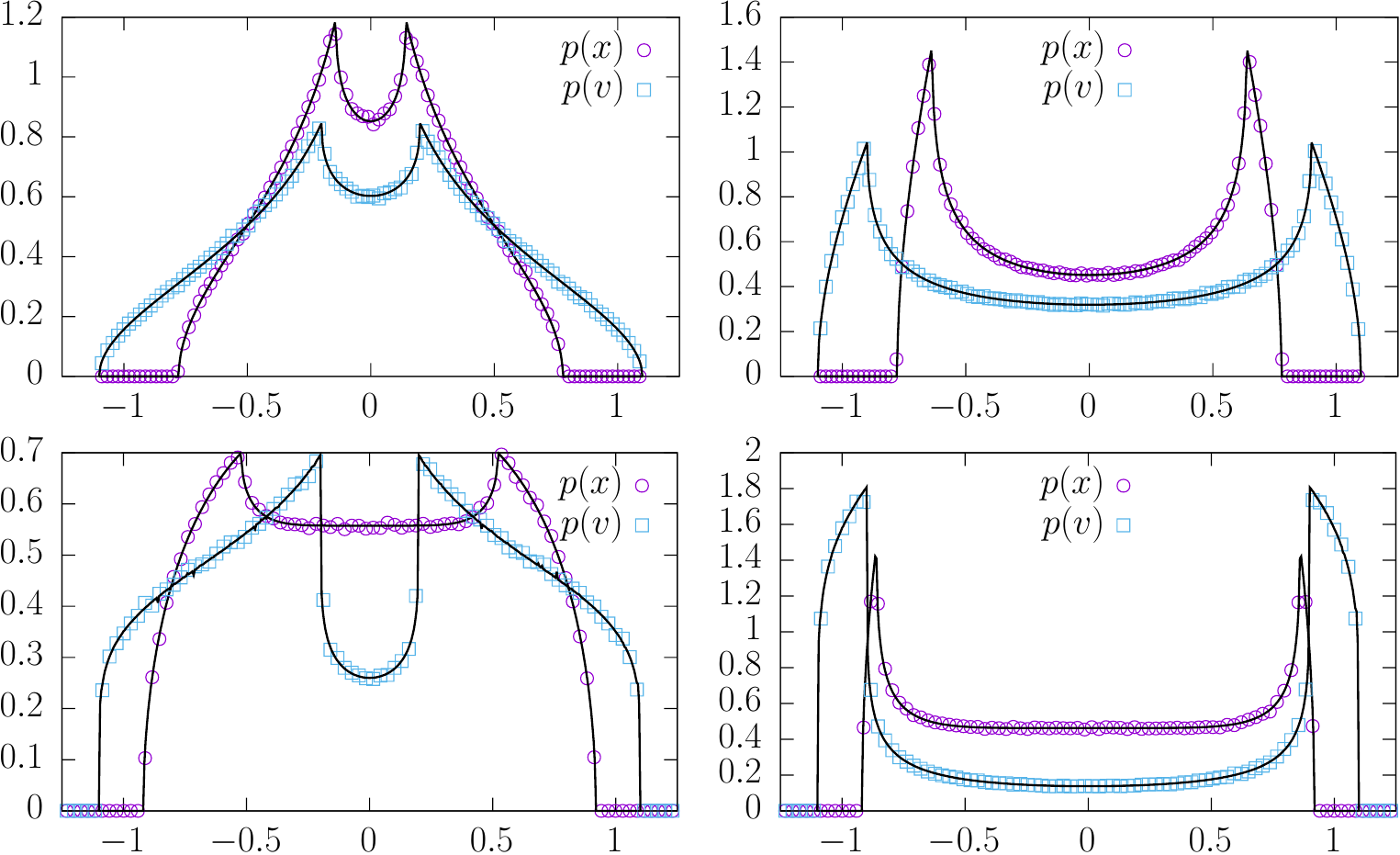}
        \caption{Stationary densities $p(x)$ and $p(v)$ for $n=2$ with $\kappa=1$ (top panel) and $n=6$ (bottom panel). The initial velocity is uniformly distributed over $[0.2,1.1]$ (left column) or $[0.9,1.1]$ (right column) intervals. Solid lines present exact results while points correspond to results of stochastic simulations of the model with velocity reversals.}
    \label{fig:ric-v}
\end{figure}

\begin{figure}
    \centering
    \begin{tabular}{c}
    \includegraphics[width=0.85\columnwidth]{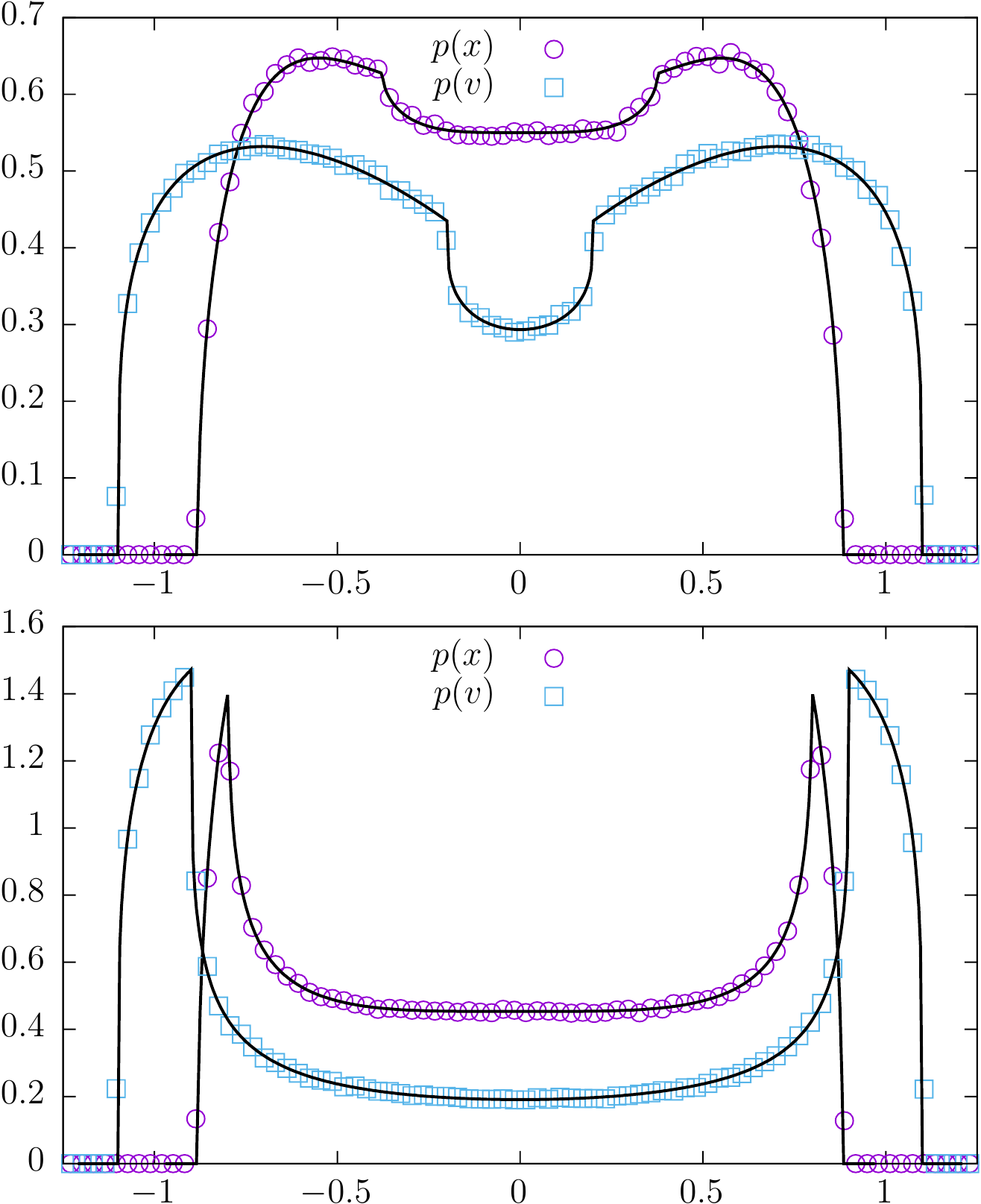}
    \end{tabular}
    \caption{Stationary densities $p(x)$ and $p(v)$ for $n=4$ with $\kappa=1$. Initial energy is uniformly distributed over $[0.2^2/2,1.1^2/2]$ (top row) $[0.9^2/2,1.1^2/2]$ (bottom row) intervals. Solid lines present exact results while points correspond to results of stochastic simulations.
}
    \label{fig:ric-e}
\end{figure}

Already for the uniform $p(\lambda)$ resulting formulas are complicated.
Consequently, we present only a few representative figures and refer the reader to~\ref{app:uniformE} for the full expressions.
Figures~\ref{fig:ric-v} and~\ref{fig:ric-e} present sample $p(x)$ and $p(v)$ distributions for random initial conditions.
In Fig.~\ref{fig:ric-v} the initial velocity is uniformly distributed over $[0.2,1.1]$ (left column) or  $[0.9,1.1]$ (right column). Various rows correspond to different potentials $V(x)=x^2$ (top row) and $V(x)=x^6$ (bottom row).
In Fig.~\ref{fig:ric-e}, the potential is fixed to $V(x)=x^4$ and the randomized parameter is the initial total energy $E$ which is uniformly distributed over $[0.2^2/2,1.1^2/2]$ (top panel)  $[0.9^2/2,1.1^2/2]$ (bottom panel). The selected support of energy distributions corresponds to minimal and maximal velocities considered in Fig.~\ref{fig:ric-v}.
Stationary densities corresponding to random initial conditions differ from the arcsine density corresponding to the fixed initial condition, see Fig.~\ref{fig:fixed-n6}. For more complicated distributions $p(\lambda)$ stationary densities have to be calculated numerically.

% \clearpage

%%%%%%%%%%%%%%%%%%%%%%%%%%%%%%%%%%%%%%%%%%%%%%%%%%%%%%%%%%%%%%%%%%%%%%%%%%%%%%%%%%%%%%%%%
\subsection{2D deterministic and randomized motions \label{sec:2d}}

\subsubsection{Deterministic motion}

The Newton equation considered in Sec.~\ref{sec:1d} can be readily generalized to a higher number of dimensions
\begin{equation}
    m \frac{d^2 \vec{r}(t)}{dt^2}=-\nabla   V(r).
    \label{eq:newton2d}
\end{equation}
Here we are interested in the case where $V(r)$ is a 2D, static single-well potential.
As exemplary potentials,
we use $V(r)=r^n/n$ with $n\in\{2,4\}$.
Trajectories of Eq.~(\ref{eq:newton2d})  are bounded, i.e., they are located within a circle of finite radius $r_{\mathrm{max}}$.
More precisely, they are located within the annulus defined by $r_{\mathrm{min}}$ and $r_{\mathrm{max}}$ radii.
As in 1D, the system is conservative, i.e., its energy is conserved, and the motion is continued along the constant energy curve.
Moreover, in the absence of velocity reversals also the angular momentum $L$ is constant.
According to the Bertrand theorem \cite{goldstein2002classical}, Eq.~(\ref{eq:newton2d}) has closed  orbits for $n=2$ (2D harmonic oscillator) and  $n=-1$ (Kepler/Coulomb problem, not considered here) \cite{arnold2010mathematical}.
In the case of $n=2$, the marginal probability distributions can be found analytically.
The marginal densities in the Cartesian coordinates,
are given by
\begin{equation}
    p_x(x)= \frac{1}{ \pi \sqrt{A_x^2-x^2}},
    \label{eq:marginal-px}
\end{equation}
where $A_x$ is the amplitude of motion along the $x$-axis.
Using the transformation of variables and Eq.~(\ref{eq:marginal-constraint}) one gets
\begin{equation}
    p_v(v_x) = \frac{1}{\pi\sqrt{(A_x \omega)^2-v_x^2}}.
    \label{eq:marginal-pv}
\end{equation}
Formulas for $p_y(y)$ and $p_v(v_y)$ are given by Eqs.~(\ref{eq:marginal-px}) and~(\ref{eq:marginal-pv}) with $x$ replaced by $y$.
The derivation of Eqs.~(\ref{eq:marginal-px}) and~(\ref{eq:marginal-pv}) is included in~\ref{app:derivation-marginal-cart}.
Analogously, marginal densities can also be found in the spherical coordinates; appropriate formulas are included in~\ref{app:derivation-spherical}.
Contrary to $n=2$, for $n=4$, solutions of Eq.~(\ref{eq:newton2d}) do not produce closed, non-circular, orbits.
Moreover, there is no exact solution for $\vec{r}(t)$.
Therefore, the approach used in the case of $n=2$ in appendices needs to be modified.
In all cases, full and marginal densities can be easily obtained numerically, and some exact results can be constructed even for $n\neq 2$.

\begin{figure}
    \centering
%     \includegraphics[width=0.6\columnwidth]{n2-2dhis_dtm3}
%     \begin{tabular}{cc}
%          \includegraphics[width=0.48\columnwidth]{n2-xhis} &
%          \includegraphics[width=0.48\columnwidth]{n2-yhis} \\
%          %
%          \includegraphics[width=0.48\columnwidth]{n2-rhis} &
%          \includegraphics[width=0.48\columnwidth]{n2-phis} \\
%     \end{tabular}
    \includegraphics[width=0.985\columnwidth]{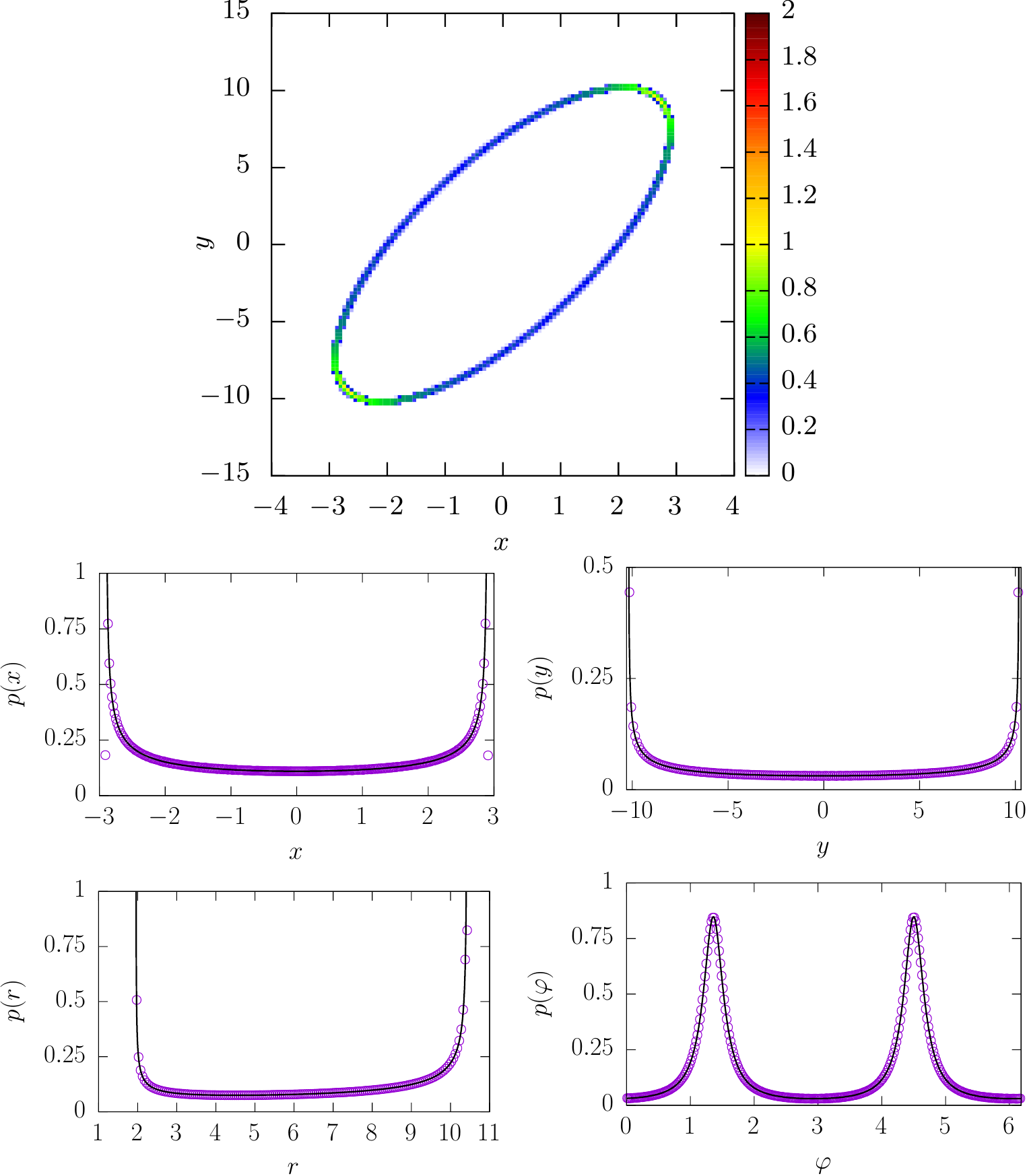}
    \caption{The $p(x,y)$ density (top row) together with marginal densities $p_x(x)$ and $p_y(y)$ (middle row), $p(r)$ and $p(\varphi)$ (bottom row).
    Solid lines in the middle and bottom row correspond to theoretical formulas.}
    \label{fig:n2}
\end{figure}

For $n=2$ the motion is performed along an ellipse.
Therefore, the probability distribution $p(x,y)$ is defined on this ellipse only. The same positions are attained at equidistant times, which for the selected parameters differ by $T=2\pi$.
The $p(x,v)$ density is shown in the top panel of Fig.~\ref{fig:n2}.
In Fig.~\ref{fig:n2} we use $r(0)=2,\dot{r}(0)=2.1,\varphi(0)=0$ and $\dot{\varphi}(0)=5.1$.
The points of maximal probability are the same as predicted by the 2nd Kepler's law.
The probability density $p(x,y)$ is maximal in places where the velocity $v(x,y)$ is minimal, i.e., in the points which are farthest from the origin (center of the ellipse).
The lower panels of Fig.~\ref{fig:n2} depict marginal densities $p_x(x)$ and $p_x(y)$ (middle row), $p(r)$ and $p(\varphi)$ (bottom row).
Solid lines present theoretical formulas for $n=2$, see Eqs.~(\ref{eq:marginal-px}), (\ref{eq:dot-varphi2}) and~(\ref{eq:general-pr}), which match results of computer simulations.
The fingerprints of the 2nd Kepler's law are also visible in $p(r)$ and $p(\varphi)$ distributions.
The maxima of $p(\varphi)$ are located at points where $\dot{\varphi}$ is minimal.
The angular velocity $\dot{\varphi}$ assumes the smallest values at the points which are the farthest from the origin (apocenter), i.e., at points corresponding to the maximal value of $r(t)$.
By analogy: minimum of $p(\varphi)$ is located at the point where angular velocity is maximal, i.e., at the point which is the closest (pericenter) to the origin.
Using the property $p(\varphi) \propto 1/\dot{\varphi}$, see Eq.~(\ref{eq:dot-varphi2}), the distribution $p(\varphi)$ is constructed in~\ref{app:derivation-spherical}.
The numerically estimated $p(\varphi)$ perfectly follows the theoretical predictions, see the right bottom panel of Fig.~\ref{fig:n2}.
Using Kepler's law, it is also possible to predict the location of extrema of $p(r)$.
Maxima of $p(r)$ are in places where $\dot{r}$ is minimal because $p(r)\propto 1/\dot{r}$, see Eq.~(\ref{eq:pr}).
The radial velocity $\dot{r}$ is minimal ($\dot{r}=0$) at points which are farthest (apocenter) and closest (pericenter) to the origin.
The $p(r)$ distribution is not symmetric, because at the apocenter the radial velocity changes faster than at the pericenter.
Consequently, for both distributions $p(\varphi)$ and $p(r)$ most and least probable values of $r$ and $\varphi$ are defined by geometric properties of the orbit, i.e., by its apocenter and pericenter.
At the same time, $p_x(x)$ and $p_y(y)$ distributions are of the arcsine type, see Eq.~(\ref{eq:marginal-px})  and~\ref{app:derivation-marginal-cart}.
These distributions are symmetric along the $x=0$ ($y=0$) line.
Here, again results of computer simulations perfectly follow theoretical densities, see the middle panel of Fig.~\ref{fig:n2}.

\begin{figure}
    \centering
%         \begin{tabular}{cc}
%         \includegraphics[width=0.44\columnwidth]{n4-trj_dtm3} &
%          \includegraphics[width=0.50\columnwidth]{n4-2dhis_dtm3} \\
%          \includegraphics[width=0.48\columnwidth]{n4-xhis} &
%          \includegraphics[width=0.48\columnwidth]{n4-yhis} \\
%          %
%          \includegraphics[width=0.48\columnwidth]{n4-rhis} &
%          \includegraphics[width=0.48\columnwidth]{n4-phis} \\
%     \end{tabular}
    \includegraphics[width=0.985\columnwidth]{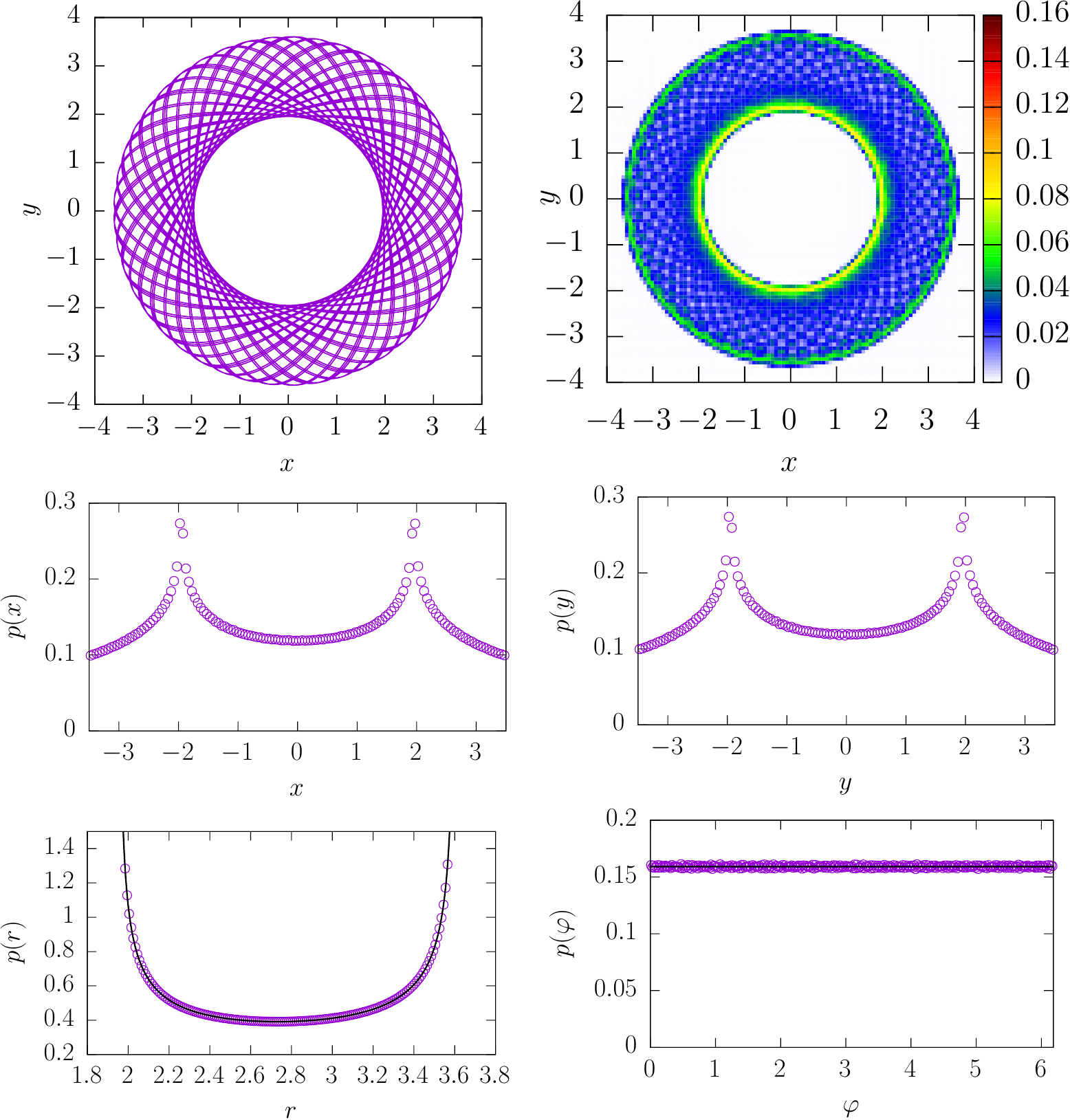}
    \caption{A sample short trajectory and the $p(x,y)$ density (top row) together with marginal densities $p_x(x)$ and $p_y(y)$ (middle row), $p(r)$ and $p(\varphi)$ (bottom row).
    Solid lines in the bottom row correspond to theoretical formulas.}
    \label{fig:n4}
\end{figure}

The results for the quartic oscillator $V(r)=r^4/4$ with the same initial conditions as for $n=2$ in Fig.~\ref{fig:n2} are presented in Fig.~\ref{fig:n4}.
Top row presents a short, 2D trajectory and 2D $p(x,y)$ density while lower panels depict marginal densities $p_x(x)$ and $p_y(y)$ (middle row), $p(r)$ and $p(\varphi)$ (bottom row).
Solid lines in the bottom row present theoretical formulas, see Eqs.~(\ref{eq:pr}) and~(\ref{eq:general-pr}), which confirm the computer simulations.
Marginal densities $p_x(x)$ and $p_y(y)$, see the middle row, are not only the same but they are symmetric along $x=0$ ($y=0$) line.

For $n=4$, although the discussion of location of maximal values of $p(r)$ and $p(\varphi)$ distributions is the same as for $n=2$, the problem becomes more complex.
On the one hand, Kepler's law still holds.
On the other hand, the orbit is no longer closed, see the left top panel of Fig.~\ref{fig:n4}.
Distances from the origin to apocenters ($r_{\mathrm{min}}$) and pericenters ($r_{\mathrm{max}}$) are fixed and well-defined, but they regularly change their location on appropriate circles.
Furthermore, the time interval between consecutive visits to the apocenter ($r_{\mathrm{min}}$) and the pericenter  ($r_{\mathrm{min}}$) is not fixed, but varies.
Consequently, it is not possible to construct $p(r)$ and $p(\varphi)$ densities in an analogous way like for $n=2$.
Nevertheless, $p(r)$ can be calculated using more general approach, see~\ref{app:derivation-spherical} where further differences between $p(r)$ and $p(\varphi)$ distributions are discussed. The most general formula for $p(r)$, see Eq.~(\ref{eq:general-pr}), reads
\begin{equation}
    p(r) \propto \frac{\sqrt{m}}{2\left[ \sqrt{E-U_{\mathrm{eff}}(r)} \right]},
    \label{eq:general-pr-main-text}
\end{equation}
where $U_{\mathrm{eff}}(r)$ is the effective potential
\begin{equation}
 U_{\mathrm{eff}}(r) = \frac{L^2}{2m r^2} + \kappa r^n,
\end{equation}
and $L$ is the angular momentum.
The system's total energy $E$ and the angular momentum $L$ are determined by the initial condition, see~\ref{app:derivation-spherical}.
Solid lines in left bottom panels of Figs.~\ref{fig:n2} and~\ref{fig:n4} correspond to Eq.~(\ref{eq:general-pr-main-text}).
It is also possible to derive analytically the $p(\dot{\varphi})$ distribution, see Eq.~(\ref{eq:general-pv}).
In this case, results of computer simulations nicely follow Eq.~(\ref{eq:general-pv}) (results not shown).

The top right panel of Fig.~\ref{fig:n4} presents the 2D histogram for the motion in the quartic potential.
The presented histogram is constructed from the long yet finite trajectory.
Within the histogram fingerprints of the deterministic trajectory are visible.
Nevertheless, already a finite realization allows for perfect reconstruction of the marginal $p(r)$ and $p(\varphi)$ densities.
With the increasing trajectory length, the internal part of the histogram becomes filled, but will not become uniform.
If $p(x,y)$ was uniform on the disk $r_{\mathrm{min}} \leqslant r \leqslant r_{\mathrm{max}}$ the $p(r)$ density would be linear.
This is definitely not the case, see the left bottom panel of Fig.~\ref{fig:n4} and Eq.~(\ref{eq:general-pr-main-text}).
Therefore, even in the $t\to\infty$ limit $p(x,y)$ will not be uniform.

\subsubsection{Randomized motion}

\begin{figure}[!h]%
\centering
% \begin{tabular}{cc}
% \includegraphics[angle=0,width=0.45\columnwidth]{n2-xhis_dtm3} &  \includegraphics[angle=0,width=0.45\columnwidth]{n2-yhis_dtm3}\\
% \includegraphics[angle=0,width=0.45\columnwidth]{n2-rhis_dtm3} &  \includegraphics[angle=0,width=0.45\columnwidth]{n2-phasehis_dtm3}\\
% \end{tabular}
\includegraphics[width=0.985\columnwidth]{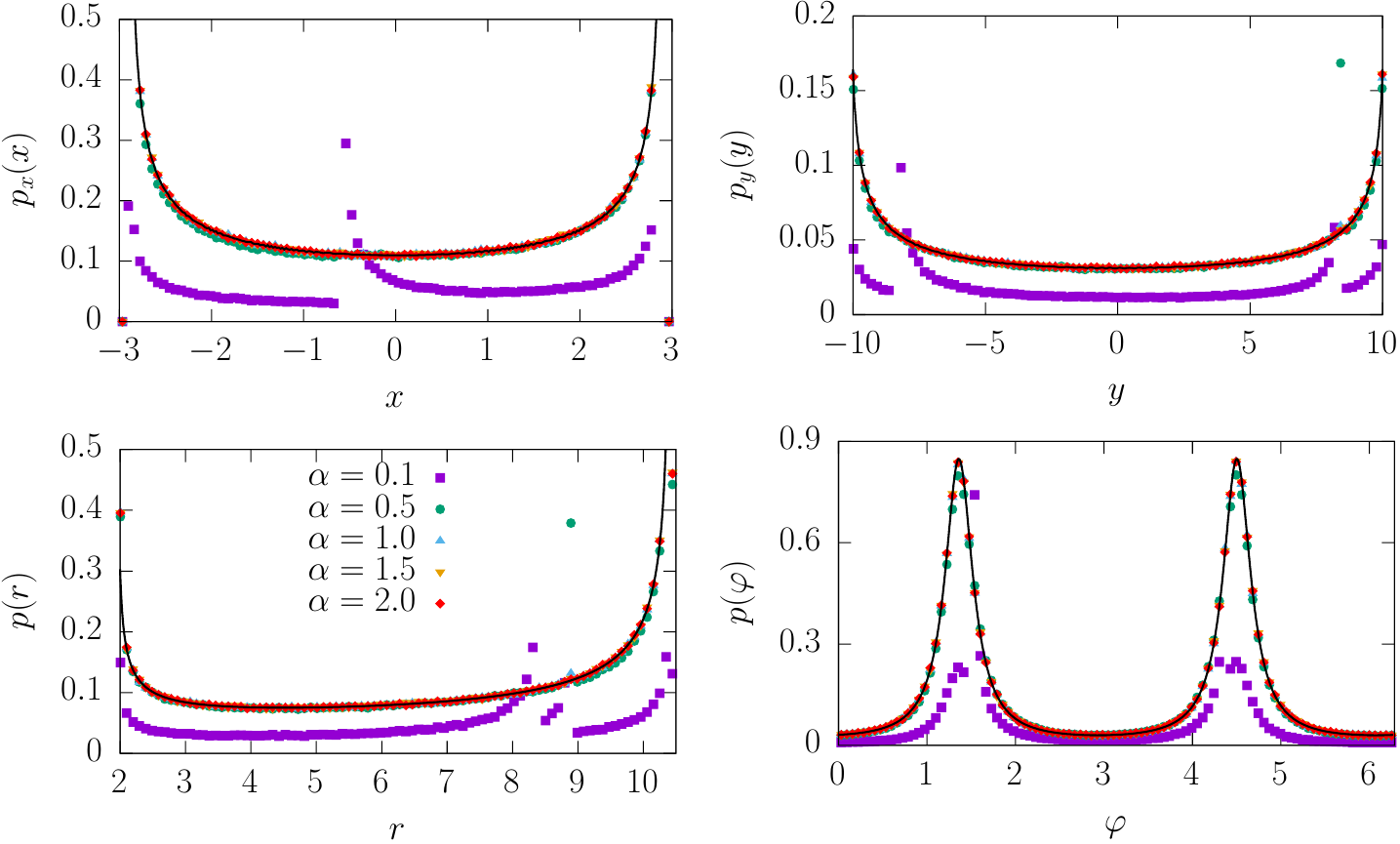}
\caption{Results of MC simulations for $n=2$. Histograms are calculated at $t=1000$.
The legend is included in the left bottom panel.
}
\label{fig:n2-mc}
\end{figure}

\begin{figure}[!h]%
\centering
% \begin{tabular}{cc}
% \includegraphics[angle=0,width=0.45\columnwidth]{n4-xhis_dtm3} &  \includegraphics[angle=0,width=0.45\columnwidth]{n4-yhis_dtm3}\\
% \includegraphics[angle=0,width=0.45\columnwidth]{n4-rhis_dtm3} &  \includegraphics[angle=0,width=0.45\columnwidth]{n4-phasehis_dtm3}\\
% \end{tabular}
\includegraphics[width=0.985\columnwidth]{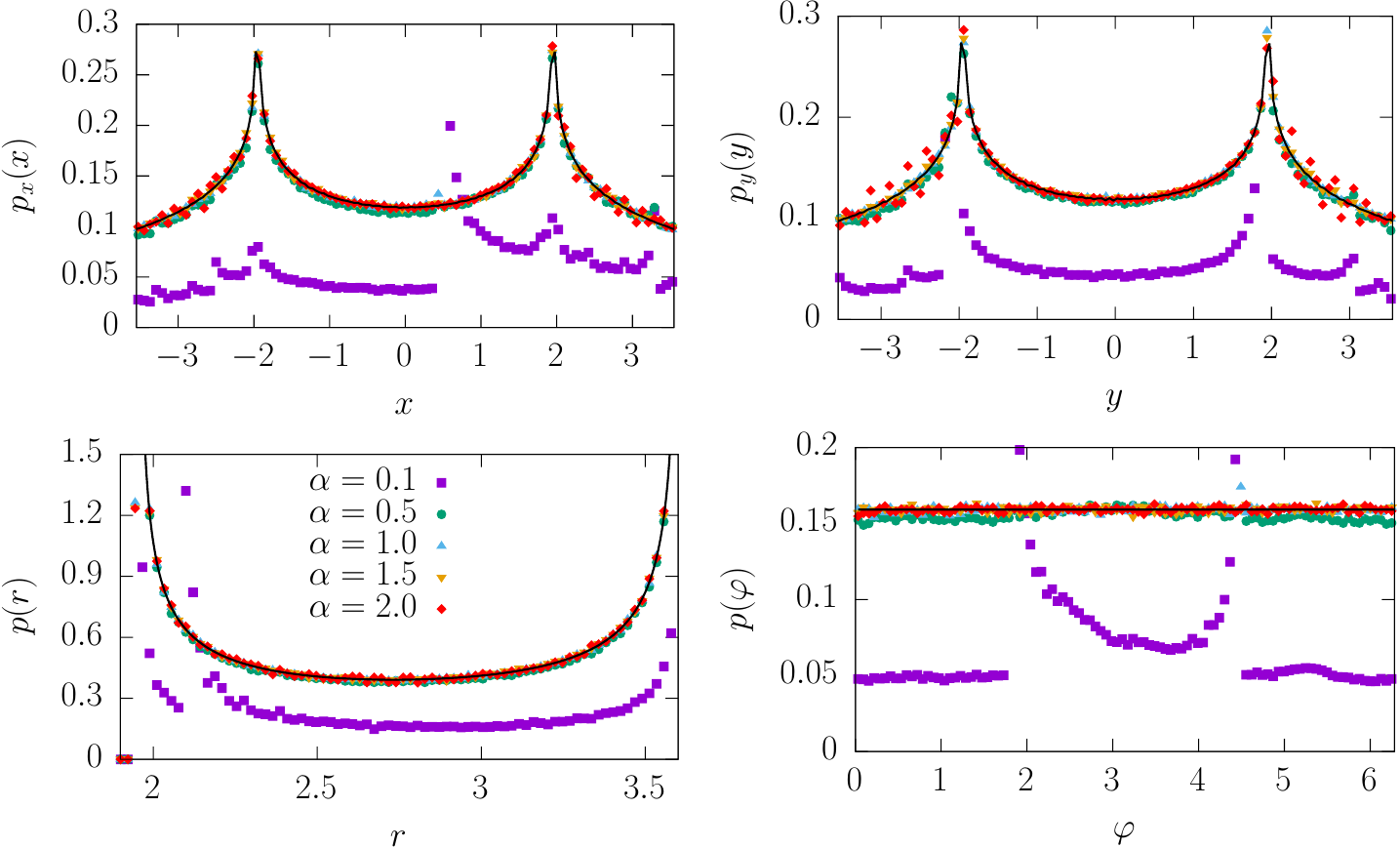}
\caption{The same as in Fig.~\ref{fig:n2-mc} for $n=4.$}
\label{fig:n4-mc}
\end{figure}

Similarly to the 1D case, the motion defined by Eq.~(\ref{eq:newton2d})  is performed along the constant energy curve which for $n\neq 2$ (in general) is not closed \cite{goldstein2002classical}.
Hard velocity reversals not only correspond to jumps in the phase space but also to change in the sign of the angular momentum $L$.
Subsequent Figs.~\ref{fig:n2-mc} and~\ref{fig:n4-mc}
depict main results of the analysis of the deterministic motion in 2D single-well potential of $r^2/2$ and $r^4/4$ type accompanied with hard velocity reversals.
Velocity reversals are performed at random times which are the sum of random increments $\tau_i$ distributed according to a heavy-tailed probability density (the same protocol as in 1D).
Analogously like in Dybiec et al. \cite{dybiec2018conservative}, we assume that $\tau_i=|\zeta_i|$, where $\zeta_i$ are independent identically distributed random variables following symmetric $\alpha$-stable density with the scale parameter equal to unity and the characteristic function given by
\begin{equation}
    \phi(k)=\exp[-|k|^\alpha].
\end{equation}
The stability index $\alpha$ ($0<\alpha\leqslant 2$) describes power-law asymptotics of symmetric $\alpha$-stable densities.
For large values of argument these densities decay as $|\zeta|^{-(\alpha+1)}$.
Consequently, for $0<\alpha<1$ they are characterized by the diverging mean, while for $1<\alpha<2$ the variance diverges.
Finally, in the limit of $\alpha=2$, they become normal distributions with all moments finite.

The 2D extension of the 1D model studied in Dybiec et al. \cite{dybiec2018conservative} displays similar properties to the 1D model.
For finite time and $\alpha$ small enough, $p_x(x)$, $p_y(y)$, $p(r)$ and $p(v)$ densities do not attain their stationary limits, which analogously like in 1D are robust with respect to the type of $p(\tau)$.
This is clearly visible for $\alpha=0.1$ and $t=1000$ where violet squares in Figs.~\ref{fig:n2-mc} and~\ref{fig:n4-mc} are placed significantly below other curves because a substantial part of the probability mass is located in the peak corresponding to the fully deterministic dynamics performed without velocity reversals.
Also for $\alpha=1$ (green bullets) some peaks are visible.
Finally, for $\alpha>1$ the mean value $\langle \tau \rangle$ is finite and densities attain their stationary shapes, which are the same as for the fully deterministic motions studied in Figs.~\ref{fig:n2} and~\ref{fig:n4}.

% \clearpage

%%%%%%%%%%%%%%%%%%%%%%%%%%%%%%%%%%%%%%%%%%%%%%%%%%%%%%%%%%%%%%%%%%%%%%%%%%%%%%%%%%%%%%%%%
\section{Summary and conclusions \label{sec:summary}}

The deterministic, frictionless, Newtonian motion in single-well 1D or 2D potentials is a universally studied example of a conservative system.
Both in 1D and 2D systems the total energy is conserved, whereas in 2D the angular momentum is also constant.
Very long observation of such a motion allows one to estimate the probability of recording a given velocity or position.

The deterministic motion can be disturbed in numerous ways; usually instantaneously breaking both momentum and energy conservation. In contrast, the studied here types of randomization preserve the energy conservation but break the angular momentum conservation.
Example of such randomization is the random velocity reversal process \cite{dybiec2018conservative}.
Despite fixed initial conditions due to velocity reversals the averaging over a single long trajectory can be used interchangeably with the ensemble averaging.
Accordingly, in 1D, it was possible to significantly extend results of Ref.~\cite{dybiec2018conservative}.
Using more general, fully analytical approach, we have obtained general analytical formulas for $p(x)$ and $p(v)$ densities, which are valid for all single-well potential of $|x|^n$ type.
In 1D stationary densities $p(x)$ and $p(v)$ are symmetric and $u$-shaped.
The deviations from general $u$-shaped densities can be introduced by the additional randomization of the initial conditions, such randomizations are also traceable analytically (for some simple cases).

Finally, the 1D model has been extended to 2D setups.
In contrast to 1D single-well potentials, 2D single-well potentials result in non-closed orbits (except for $n=2$) and quasi-periodic motion. Nevertheless, even in such a case, it is possible to derive exact formulas, at least for marginal $p_x(x),p_y(y),p_v(\dot{x}),p_v(\dot{y})$ and $p(r)$ densities which asymptotically are insensitive to velocity reversals.

%%%%%%%%%%%%%%%%%%%%%%%%%%%%%%%%%%%%%%%%%%%%%%%%%%%%%%%%%%%%%%%%%%%%%%%%%%%%%%%%%%%%%%%%%
%%
%% ACKNOWLEDGMENTS
%%

% \begin{acknowledgments}

\ack

This research was supported in part by PLGrid Infrastructure.
Computer simulations have been performed at the Academic
Computer Center Cyfronet, AGH University of Science and Technology (Krak\'ow, Poland)
under CPU grant ``DynStoch''.
Discussions with Ralf Metzler and Aleksei Chechkin are greatly acknowledged.

% \end{acknowledgments}

%\bibliography{core-bibliography}

%%%%%%%%%%%%%%%%%%%%%%%%%%%%%%%%%%%%%%%%%%%%%%%%%%%%%%%%%%%%%%%%%%%%%%%%%%%%%%%%%%%%%%%%%
%%
%% appendix
%%
\appendix

%%%%%%%%%%%%%%%%%%%%%%%%%%%%%%%%%%%%%%%%%%%%%%%
\section{Derivation of formula for $p(x)$ and $p(v)$ \label{app:derivation-p}}

In Dybiec et. al.~\cite{dybiec2018conservative} formulas for $p(x)$ and $p(v)$ have been derived using semi-analytical and numerical methods for the potential of $|x|^n$ type with $n\in\{1,2,4,\infty\}$.
Here, we present general, fully analytical, derivation that is valid for any exponent $n$.

For the 1D periodic motion the probability of finding a particle at the point $x$ is proportional to the time spent in the vicinity of this point
\begin{eqnarray}
    p(x)dx  & = &  \frac{dt}{\frac{T}{2}} = \frac{2}{T}  \frac{dt}{dx}dx = \frac{2}{T}\frac{dx}{\frac{dx}{dt}} = \frac{2}{T}\frac{dx}{v},
    \label{eq:dtration}
\end{eqnarray}
where $T$ is the period of the motion \cite{landau1988theoretical} in $V(x)=\kappa |x|^n$ ($\kappa>0,n>0$)
\begin{equation}
    T=\frac{2}{n}\sqrt{\frac{2\pi m}{E}}\left[ \frac{E}{\kappa } \right]^{\frac{1}{n}} \frac{\Gamma\left( \frac{1}{n} \right)}{\Gamma\left( \frac{1}{2} + \frac{1}{n} \right)}.
    \end{equation}
The velocity $v=\dot{x}$ can be calculated from the energy conservation principle
\begin{equation}
    E=\frac{1}{2} m v^2 + V(x) = \frac{1}{2} mv^2+\kappa  |x|^n
\end{equation}
resulting in
\begin{equation}
    v=\sqrt{2\left[ E-\kappa  |x|^n  \right]}.
\end{equation}
The total energy $E$ is determined by the initial condition $E=\frac{1}{2} mv_0^2+\kappa  |x_0|^n$.
Finally, the formula for the $p(x)$ density reads
\begin{equation}
    p(x)= \frac{2}{T}\frac{1}{\sqrt{2\left[ E-\kappa  |x|^n  \right]}}.
    \label{eq:px-app}
\end{equation}
The particle can be most easily found at points where it spends a long time, see Eq.~(\ref{eq:dtration}), i.e., at points where the velocity is minimal.
These points correspond to places at which the velocity is softly reversed, i.e., to the points which are in the maximal distance from the origin, see Eq.~(\ref{eq:px-app}).

The $p(v)$ density  can be obtained by the transformation of variables
\begin{equation}
    p(v)=p(x(v)) \left| \frac{dx}{dv} \right|.
    \label{eq:trans}
\end{equation}
From the energy conservation
\begin{equation}
    x= \left[ \frac{1}{\kappa } \left(  E-\frac{1}{2}mv^2  \right)  \right]^{\frac{1}{n}}
\end{equation}
and
\begin{equation}
    \frac{dx}{dv}=- \frac{m}{\kappa n}\left[ \frac{1}{\kappa } \left(  E-\frac{1}{2}mv^2  \right)  \right]^{\frac{1}{n}-1} v.
    \label{eq:derivative}
\end{equation}
Finally, after substituting Eq.~(\ref{eq:derivative}) in Eq.~(\ref{eq:trans}) and using the fact that $E-\kappa |x|^n=\frac{1}{2}mv^2$ one obtains
\begin{equation}
    p(v)=\frac{2m}{T\kappa n} \left[ \frac{1}{\kappa } \left( E - \frac{1}{2}m v^2 \right)  \right]^{\frac{1}{n}-1}.
\end{equation}
Maxima of the $p(v)$ density are located at points where the acceleration is minimal, i.e., the velocity is maximal.
At the origin ($x=0$), the velocity is maximal while the acceleration is minimal.
Consequently, at the origin, the $p(x)$ is minimal, while $p(v)$ is maximal.

%%%%%%%%%%%%%%%%%%%%%%%%%%%%%%%%%%%%%%%%%%%%%%%
\section{Uniform energy distribution\label{app:uniformE}}

In simplest cases formulas~(\ref{eq:comp-x}) and~(\ref{eq:comp-v}) can be integrated analytically.
For instance, if the randomized parameter is the total energy $E$ and it is uniformly distributed over $[E_1,E_2]$ one can calculate $\tilde{p}(E,x)=\int p(x|E)p(E)dE$ and $\tilde{p}(E,v)=\int p(v|E)p(E)dE$ resulting in
\small
\[
\hspace{-2cm}\tilde{p}(E,x)=
 \frac{E n \sqrt{\frac{m}{E}} \Gamma \left(\frac{1}{2}+\frac{1}{n}\right) \left(\frac{E}{\kappa}\right)^{-1/n} \left(\frac{E \left| x\right| ^{-n}}{\kappa}\right)^{\frac{1}{n}-\frac{1}{2}} \, _2F_1\left(\frac{1}{2},\frac{1}{n}-\frac{1}{2};\frac{3}{2};1-\frac{E \left| x\right| ^{-n}}{\kappa}\right) \sqrt{\frac{E-\kappa \left| x\right| ^n}{m}}}{\sqrt{\pi } \Gamma \left(\frac{1}{n}\right)}
\]
\normalsize
for $E-\kappa \left| x\right| ^n \geqslant 0$, otherwise $\tilde{p}(E,x)=0$.
Analogously
\begin{eqnarray}
\hspace{-2cm} \tilde{p}(E,v)  = &&
 -\frac{\sqrt{\frac{2}{\pi }} n \sqrt{E m} \Gamma \left(\frac{1}{2}+\frac{1}{n}\right) \left(1-\frac{m x^2}{2 E}\right)^{1/n} \left(1-\frac{2 E}{m x^2}\right)^{-1/n} }{(n-2) \Gamma \left(\frac{1}{n}\right)} \\
 && \times  \left[\, _2F_1\left(\frac{n-2}{2n},\frac{n-1}{n};\frac{3n-2}{2n};\frac{2 E}{m x^2}\right)-\, _2F_1\left(\frac{n-2}{2n},-\frac{1}{n};\frac{3n-2}{2n};\frac{2 E}{m x^2}\right)\right] \nonumber
\end{eqnarray}
\normalsize
for $E-\frac{m}{2} v^2 \geqslant 0$, otherwise $\tilde{p}(E,v)=0$.

Finally, the stationary distributions $p(x)$ and $p(v)$ read
\begin{equation}
p(x)=\frac{\tilde{p}(E_2,x)-\tilde{p}(E_1,x)}{E_2-E_1}
\end{equation}
and
\begin{equation}
p(v)=\frac{\tilde{p}(E_2,v)-\tilde{p}(E_1,v)}{E_2-E_1}.
\end{equation}

%%%%%%%%%%%%%%%%%%%%%%%%%%%%%%%%%%%%%%%%%%%%%%%
\section{Motion in $V(r)=\kappa r^2$  \label{app:derivation-marginal-cart}}

For $V(r)=\kappa r^2$  ($\kappa>0$) orbits of the motion in 2D are closed.
Moreover, in the Cartesian coordinates, Eq.~(\ref{eq:newton2d}) separates into two independent equations
\begin{equation}
    \left\{
    \begin{array}{lcl}
    m\ddot{x}(t) & = & -2\kappa  x(t) \\
    m\ddot{y}(t) & = & -2\kappa  y(t) \\
    \end{array}
    \right.,
\end{equation}
which can be easily solved.
For instance, the formula for $x(t)$ reads
\begin{equation}
    x(t)=A_x \cos(\omega t + \Delta_x),
\end{equation}
where $\omega^2=2\kappa/m$, $\Delta_x = - \arctan{\frac{\dot{x}_0}{\omega x_0}}$ and $A_x=\sqrt{x_0^2+\dot{x}^2_0/\omega^2}$.
The analogous formula holds for $y(t)$.
The marginal density $p_x(x)$ can be found in an analogous way like $p(x)$ in~\ref{app:derivation-p}
\begin{equation}
    p_x(x)dx = \frac{2}{T} \frac{dx}{\dot{x}}.
\end{equation}
The velocity $\dot{x}$ as a function of the position $x$ can be calculated from the condition
\begin{equation}
    \left[ \frac{x(t)}{A_x} \right]^2+ \left[ \frac{\dot{x}(t)}{A_x \omega} \right]^2=1,
    \label{eq:marginal-constraint}
\end{equation}
which allows to calculate
\begin{equation}
    p_x(x)=\frac{2}{T\omega \sqrt{A_x^2-x^2}} = \frac{1}{ \pi \sqrt{A_x^2-x^2}}.
    \label{eq:marginal-px-app}
\end{equation}
Using the transformation of variables and Eq.~(\ref{eq:marginal-constraint}) one gets
\begin{equation}
    p_v(\dot{x}=v) = p_x(x) \left|\frac{dx}{dv} \right|=\frac{1}{\pi\sqrt{(A_x \omega)^2-v^2}}.
    \label{eq:marginal-pv-app}
\end{equation}
Marginal densities $p_y(y)$ and $p_v(\dot{y}=v)$ are given by Eqs.~(\ref{eq:marginal-px-app}) and~(\ref{eq:marginal-pv-app}) with $x$ replaced by $y$, i.e., $A_x \to A_y$, $x \to y$ and $\dot{x} \to \dot{y}$.
These densities are arcsine-like and even.

%%%%%%%%%%%%%%%%%%%%%%%%%%%%%%%%%%%%%%%%%%%%%%%
\section{2D Motion in $V(r)=\kappa r^n$  \label{app:derivation-spherical}}
The equation of motion (\ref{eq:newton2d}) in the central potential  $V(r)=\kappa r^n$  ($\kappa>0$ and $n>0$) can be rewritten in the polar coordinates $(r,\varphi)$ as
\begin{equation}
\label{eq:eff2d}
\left\{
\begin{array}{lcl}
m\ddot{r}(t) & = & -U_{\mathrm{eff}}'(r) \\
     L & = & \mathrm{const}  \\
     \end{array}
     \right.,
\end{equation}
where $U_{\mathrm{eff}}(r)$ is the effective potential
\begin{equation}
    U_{\mathrm{eff}}(r) = \frac{L^2}{2m r^2} + \kappa r^n,
    \label{eq:effective-potential}
\end{equation}
and $L$ is the angular momentum
\begin{equation}
    L=m r^2(t) \dot{\varphi}(t).
\end{equation}
The angular momentum is conserved and its value is determined by the initial condition
\begin{equation}
    L = m \vec{r}_0 \times \dot{\vec{r}}_0=  m r_0^2\dot{\varphi}_0.
\end{equation}
For $n=2$ and the special case of $\omega^2= 2\kappa/m=1$, $m=1$ the solution of Eq.~(\ref{eq:eff2d}) can be found analytically
\begin{equation}
r^2(t) = \frac{1}{2}\left[\left(-\frac{L^2}{r_0^2}+r_0^2-\dot{r}_0^2\right) \cos(2 t)+\frac{L^2}{r_0^2}+r_0^2+\dot{r}_0^2 + 2 r_0 \dot{r}_0 \sin(2 t)\right].
\label{eq:rt}
\end{equation}
From Eq.~(\ref{eq:rt}) and the conservation of angular momentum $L$ the angular velocity $\dot{\varphi}(t)$ together with $\varphi(t)$ and its inversion $\varphi^{-1}(t)$ can be evaluated. Here, for brevity, we present only $\varphi(t)$
\begin{equation}
\varphi(t) = \arctan \left[ \frac{\sec t \left(L^2 \sin t+\dot{r}_0^2 r_0^2 \sin t + \dot{r}_0 r_0^3 \cos t\right)}{L r_0^2}\right].
\end{equation}
Note, that initially $ \varphi(0)= \arctan{\left[ \frac{m r_0 \dot{r}_0}{L}\right]}$.
For such an initial condition, the probability distribution $p(\varphi)$ can be obtained analytically
\begin{eqnarray}
\label{eq:dot-varphi2}
     p(\varphi) d\varphi & = & \frac{1}{T}\frac{d\varphi}{\dot{\varphi}} = \frac{r^2(\varphi) d\varphi }{T L}  \\
    & = &  \frac{L r_0^2 \left(L^2+r_0^2 \dot{r}_0^2\right) \sec ^2 \varphi }{2\pi  \left(L^4+L^2 r_0^4 \tan ^2\varphi +\dot{r}_0^2 r_0^2 \left(2 L^2+r_0^4+\dot{r}_0^2 r_0^2\right)-2 L \dot{r}_0 r_0^5 \tan \varphi \right)}d\varphi. \nonumber
\end{eqnarray}
Analogously to $p(\varphi)$, the marginal density $p(r)$ can be found from
\begin{equation}
    p(r)dr \propto \frac{dr}{ \dot{r}}.
    \label{eq:pr}
\end{equation}
In Eq.~(\ref{eq:pr}) there is the $\propto$ sign.
The origin of this proportionality is visible for $n=2$.
It originates due to the fact that the same radii $r$ are taken more than once during one period of motion.
More precisely, minimal and maximal values of $r$ are recorded twice, see below, while intermediate values are taken four times.
The formula for $p(r)$ can be derived with the help of Eq.~(\ref{eq:pr}) and Eq.~(\ref{eq:rt}).
Unfortunately, it is not the most straightforward approach and the constructed formula is quite lengthy.
The more convenient approach is to utilize the energy conservation and the angular momentum conservation, see below.

For $n \neq 2$ orbits do not close; nevertheless, the motion is bounded.
The motion is restricted to  $r(t) \in [r_{\mathrm{min}},    r_{\mathrm{max}}]$ where $r_{\mathrm{min}}$ and $r_{\mathrm{max}}$ are determined by the condition
\begin{equation}
    U_{\mathrm{eff}}(r) = E,
\end{equation}
where $E$ is the total (conserved) energy and $U_{\mathrm{eff}}(r)$ is the effective potential given by Eq.~(\ref{eq:effective-potential}).
Due to boundness of the motion marginal densities $p_x(x)$ and $p_y(y)$ as well as $p_v(v_x)$ and $p_v(v_y)$ have the same support and functional dependence.
Please note however, that for $n \neq 2$, the marginal density $p(\varphi)$ cannot be constructed using the approach applied in Eq.~(\ref{eq:dot-varphi2}).
First of all, orbits are not closed; thus the period of the motion is not defined.
After increasing $\varphi(t)$ by $2\pi$ the radius $r(t)$ takes a different value, more precisely $r(\varphi)\neq r(\varphi+2\pi)$.
Therefore, due to the conservation of the angular momentum
\begin{equation}
    L = m r^2(t) \dot{\varphi}(t) = \mathrm{const},
\end{equation}
the mapping $\varphi \to \dot{\varphi}$ is not unique.
Nevertheless, the $p(\varphi)$ density can be estimated numerically.
From computer simulations it seems that $p(\varphi)$ is uniform.
At the same time the marginal density $p(r)$ still can be constructed with the help of Eq.~(\ref{eq:pr}).
Despite lack of periodicity, due to the energy and momentum conservations
\begin{equation}
    \frac{1}{2}m\dot{r}^2(t) = E -U_{\mathrm{eff}}(r),
\end{equation}
the mapping $r\to\dot{r}$ is unique and Eq.~(\ref{eq:pr}) is valid.
The formula for $p(r)$ reads
\begin{equation}
    p(r) \propto \frac{1}{\dot{r}} = \frac{\sqrt{m}}{ \sqrt{2\left[E-U_{\mathrm{eff}}(r)\right]} },
    \label{eq:general-pr}
\end{equation}
where $U_{\mathrm{eff}}(r)$ is given by Eq.~(\ref{eq:effective-potential}) and $E$ and $L$ are determined by the initial condition.
Eq.~(\ref{eq:general-pr}) gives the general formula for the $p(r)$ density for any 2D single-well potential.
Practically, the only difficulty is the calculation of the $r_{\mathrm{min}}$, $r_{\mathrm{max}}$ and the normalization constant $N$
\begin{equation}
N^{-1}=\int_{r_{\mathrm{min}}}^{r_{\mathrm{max}}} \frac{dr}{\dot{r}}.
\end{equation}
Furthermore, using the transformation of variables and the relation between the angular momentum and the radial velocity from Eq.~(\ref{eq:general-pr}) it is possible to calculate
\begin{eqnarray}
\label{eq:general-pv}
    p(\dot{\varphi})&=&p\left(r=\frac{\sqrt{L}}{\sqrt{m\dot{\varphi}}}\right)
    \frac{\sqrt{L}}{2\sqrt{m\dot{\varphi^3}}} \\ \nonumber
    & \propto &
    \frac{\sqrt{m}}{ \sqrt{2\left[E-U_{\mathrm{eff}}( {\sqrt{L}}/{\sqrt{m\dot{\varphi}}} )\right]} }
    \frac{\sqrt{L}}{2\sqrt{m\dot{\varphi^3}}}.
\end{eqnarray}
The support of $p(\dot{\varphi})$ density is determined by the $L=mr^2\dot{\varphi}$ condition. In accordance with Kepler's law, the minimal (maximal) radial velocity corresponds to the maximal --- $r_{\mathrm{max}}$ (minimal --- $r_{\mathrm{min}}$) radius.

\section*{References}
% \bibliography{core-bibliography}

\def\url#1{}

\end{document}